\documentclass[aps,prl,groupedaddress,twocolumn]{revtex4}
\usepackage{graphicx}  
\usepackage{epstopdf}
\usepackage{amsthm}
\usepackage{amssymb}

\begin{document}
\title{Bubble dynamics in N dimensions}
\author{Alexander R. Klotz\footnote{klotza@physics.mcgill.ca}}
\affiliation{Department of Physics, McGill University}
\begin{abstract}Cavitation and bubble dynamics are central concepts in engineering, the natural sciences, and the mathematics of fluid mechanics. Due to the nonlinear nature of their dynamics, the governing equations are not fully solvable. Here, the dynamics of a spherical bubble in an $N$-dimensional fluid are discussed in the hope that examining bubble behavior in $N$ dimensions will add insight to their behavior in three dimensions. Several canonical results in bubble dynamics are re-derived, including the Rayleigh collapse time, the Rayleigh-Plesset equation, and the Minnaert frequency. Numerical simulations are used to examine the onset of nonlinear behavior. Overall, the dynamics of bubbles are faster at higher dimensions, with nonlinear behavior occurring at lower amplitudes. Several features are found to be unique to three dimensions, including the trend of nonlinear behaviour and apparent coincidences in timescales. \end{abstract}
\maketitle

\section{Introduction}

The behavior of bubbles and cavities has been an area of interest in fluid dynamics since the time of Leonardo da Vinci \cite{ProspBub}. Serious theoretical research began in the early twentieth century when Lord Rayleigh derived their collapse time in order to study the effect of cavitation damage to ship propellers \cite{rayleigh}. Bubbles and cavitation remain an active area of research in fluid dynamics \cite{obreschkow} and soft condensed matter physics \cite{tabor}. Notably, bubbles find application as contrast agents for biomedical ultrasound, where their oscillations can be detected acoustically \cite{klotz}. The mechanism of sonoluminescence, a flash of light released by a collapsing bubble, is still an unsolved problem in physics \cite{sono}.

The dynamics of gas-filled spherical bubbles are described using the Rayleigh-Plesset equation, a special case of the Navier-Stokes equation that describes the oscillation of a spherical cavity in an infinite incompressible fluid. In the linear regime the bubble oscillates harmonically, pulsating with a breathing mode given by the characteristic Minnaert frequency. At higher oscillation amplitudes, the bubble exhibits strong nonlinear behavior. Like the Navier-Stokes equation, there is no known general solution to the Rayleigh-Plesset equation or to the Rayleigh-like collapse of a bubble. The equations are often solved numerically, but in many cases the oscillations may become chaotic.

Higher or lower dimensional analyses are often utilized to aid understanding of three dimensional physical problems. For example, the Ising model of ferromagnetism, while not solvable in three dimensions, can be exactly solved in two dimensions and is solvable with mean-field theory in more than four dimensions \cite{cipra}. Varying the dimensionality of the system as a free parameter can often provide fresh physical or mathematical insight \cite{Nbrenner}. Higher dimensional generalizations have previously been utilized to solve theoretical problems in acoustics: the question ``Can one hear the shape of a drum?" was first answered on a sixteen dimensional torus \cite{Milnor}. Regularity criteria established for the Navier-Stokes equations in $N$ dimensions \cite{Fan} have been used to establish such criteria for the three-dimensional case \cite{Zhou}. Bubble dynamics, particularly nucleation and expansion, are also of interest in high-energy physics in the areas of relativistic hydrodynamics \cite{relativistic} and cosmological phase transitions in three or more spatial dimensions \cite{string}. The stability of fluid droplets and cylinders were analyzed in $N$ dimensions by Cardoso and Gaultieri \cite{NDrop} and differential equations resembling certain types of $N$-dimensional bubble were analyzed by Lima et al. \cite{LimaBubble}, but their analysis focused on the three-dimensional case. Notably, Prosperetti discussed Rayleigh-like and oscillatory bubble dynamics in $N$ dimensions but only focused on the three-dimensional case and presented inconsistent results \cite{ProspBub}.

In this paper, the dynamics of a hyperspherical bubble are analyzed. Relevant terms such as the Rayleigh collapse time, Rayleigh-Plesset equation, and the Minnaert frequency are derived. A recent paper by Obreschkow et al. \cite{obreschkow} presenting analytical approximations to bubble collapse will also be analyzed in its generalized form. As a case study into higher dimensional bubble dynamics, the onset of nonlinear behavior is analyzed. Examining the behavior of bubbles as a function of dimensionality, it is hoped that higher dimensional bubble dynamics cast light on unknown aspects of three-dimensional bubble dynamics.

The derivations in this paper follow those historically used for three dimensions. Various assumptions are made: the bubble remains spherical at all times, the fluid is incompressible and infinite, the gas inside the bubble is an ideal gas with uniform pressure and constant number. When comparing analytical results to numerical solutions, the Runge-Kutta method was used for second-order differential equations, and the Euler method was used for first-order differential equations. 

\section{Derivations}

\subsection*{The Continuity Equation in $N$ Dimensions}

Before the dynamics of bubbles are re-derived in arbitrary dimension, it is necessary to understand the behavior of fluids, and the associated vector calculus, in arbitrary dimension. Central to this is the continuity equation that formally states the conservation of mass, relating the  velocity field $u$ and the density $\rho$ in a fluid:
\begin{equation}
\frac{\partial \rho}{\partial t}+\nabla \cdot\left(\rho u\right)=0
\label{eq:cont1}
\end{equation}
In an incompressible fluid, $\rho$ is constant and the equation is simplified:
\begin{equation}
\nabla \cdot u=0
\label{eq:cont2}
\end{equation}
In three-dimensional spherical coordinates with no angular dependence, the divergence operator takes the form:
\begin{equation}
\nabla\cdot\ u=\frac{1}{r^2}\frac{d}{dr}\left(r^{2}u\right)
\label{eq:del3}
\end{equation}
Solving the continuity equation with spherical symmetry shows that the velocity field around a spherical source has an inverse-square fall-off:
\begin{equation}
u(r)=\frac{u_{o}}{r^2}
\label{eq:rsqu}
\end{equation}
In $N$ dimensions, the spherically symmetric divergence operator takes the form:
\begin{equation}
\nabla\cdot\ u=\frac{1}{r^{N-1}}\frac{d}{dr}\left(r^{N-1}u\right)
\label{eq:delN}
\end{equation}
and the spherical solution of the continuity equation is:
\begin{equation}
u(r)=\frac{u_{o}}{r^{N-1}}
\label{eq:rsquN}
\end{equation}
This relationship will be used when deriving the kinetic energy of the fluid around a pulsating hyperspherical bubble. It is also worthwhile to express the surface area and volume of the unit sphere in $N$ dimensions:
\begin{equation}
S_{N}=\frac{2\pi^{N/2}}{\Gamma{\left(N/2\right)}}\, \ \ V_{N}=\frac{\pi^{N/2}}{\Gamma{\left(N/2+1\right)}}=N\cdot S_{N}
\label{eq:Sn}
\end{equation}

\subsection*{Rayleigh-like Collapse}

Rayleigh in 1917 \cite{rayleigh} derived a formula for the collapse time of a spherical cavity of radius $R_o$ in an infinite liquid with pressure $P$ and density $\rho$, considering the time it takes for the liquid to fill the cavity. The derivation lay in equating the kinetic energy of the fluid with the work performed by the bubble. The result was:
\begin{equation}
t_{\mathrm{collapse}}=0.91468\cdot R_{o}\sqrt{\frac{\rho}{P}}
\label{eq:RC3}
\end{equation}
The collapse time can be derived in $N$ dimensions: the kinetic energy is calculated by integrating along the radial coordinate $r$ the energies of each spherical volume element of density $\rho$ and thickness $dr$ as it expands or contracts in phase with the central bubble of radius $R$.
\begin{equation}
KE=\int_{R}^{\infty}S_{N}r^{N-1}\rho\left[\dot{R}\left(\frac{R}{r}\right)^{N-1}\right]^{2}dr=\frac{S_{N}\rho}{2}\frac{\dot{R}^{2}R^{N}}{N-2}
\label{eq:energy}
\end{equation}
The term in the square brackets represents the continuity-dependent velocity dispersion, (\ref{eq:rsquN}). In two dimensions, the kinetic energy integral does not converge; the dimensionality discussed in this paper is limited to $N \geq 3$. In the case described by Rayleigh, the pressure P in the fluid is constant during collapse. Thus, the work performed by the fluid is merely the change in volume multiplied by the pressure:
\begin{equation}
W=V_{N}P\left(R_o^{N}-R^{N}\right)
\label{eq:ConstP}
\end{equation}
Recalling that the ratio $V_{N}/S_{N}$ is simply equal to 1/N, equating (\ref{eq:ConstP}) and (\ref{eq:energy}) and isolating $\dot{R}$ yields a first-order differential equation between $R$ and $t$.
\begin{equation}
\left(\frac{dR}{dt}\right)^{2}=\frac{P}{\rho}\frac{2\left(N-2\right)}{N}\left(\frac{R_{o}^{N}}{R^{N}}-1\right)
\label{eq:drdt}
\end{equation}
The Rayleigh collapse time $t_{RC}$, the amount of time required for a bubble initially at $R_{o}$ to collapse to zero, can be found by isolating $dt$ from (\ref{eq:drdt}) and integrating by $dR$ from 0 to $R_{o}$:
\begin{equation}
t_{RC}=\sqrt{\frac{\rho}{P}\frac{N}{2\left(N-2\right)}}\int_{0}^{R_{o}}\sqrt{\frac{R^N}{R_{o}^{N}-R^N}}dR
\label{eq:trc1}
\end{equation}
Following the procedure used for $N=3$ by Rayleigh, the substitution $\beta=\frac{R}{R_o}$ is made:
\begin{equation}
t_{RC}=R_o\sqrt{\frac{\rho}{P}\frac{N}{2\left(N-2\right)}}\int_{0}^{1}\frac{\beta^{N/2}}{\left(1-\beta^N\right)^{1/2}}d\beta
\label{eq:trc2}
\end{equation}
This integral can be solved in terms of gamma functions:
\begin{equation}
t_{RC}=\sqrt{\frac{\pi}{2N\left(N-2\right)}}\frac{\Gamma\left(\frac{1}{2}+\frac{1}{N}\right)}{\Gamma\left(1+\frac{1}{N}\right)}R_o\sqrt{\frac{\rho}{P}} \equiv \tau\cdot R_o\sqrt{\frac{\rho}{P}}
\label{eq:trc}
\end{equation}
We define the unitless parameter $\tau$ to simplify notation.
For $N=3$, $\tau$ reduces to the familiar result of 0.91468, and is a monotone decreasing function as a function of $N$ for $N\geq3$. For large $N$, the gamma functions in the numerator and denominator simplify to $\sqrt{\pi}$ and 1 respectively. The collapse behavior will be discussed in greater detail subsequently, but the chief result of this derivation is that the collapse time decreases in higher dimensions.

\subsection*{Analytical Approximations}

A recent paper by Obreschkow et al. \cite{obreschkow} examined Rayleigh-like collapse in greater detail. They expressed the collapse of a spherical bubble in normalized form, and derived a simple function to approximate it. Their approximation function, even at lowest-order, showed excellent agreement with both the numerical solution of the Rayleigh equation and their experimental measurements. Here, the normalization and approximation will be generalized for higher dimensions in order to fully explore the collapsing bubble behavior.

\begin{figure}
	\centering
		\includegraphics[width=0.45\textwidth]{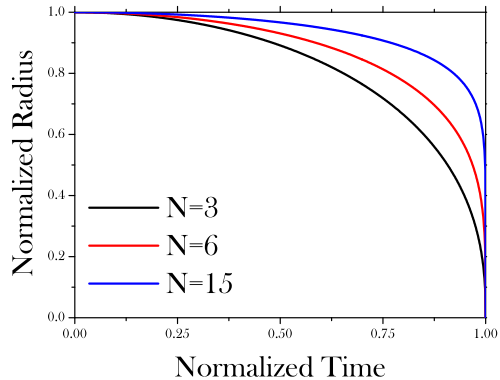}
	\caption{Normalized collapse of a bubble with non-dimensionalized radius, pressure, and density with respect to time in 3, 6, and 15 dimensions.}
	\label{fig:bub1}
\end{figure}

The Rayleigh collapse equation is normalized with the variables $r\equiv R/R_{o}$ and $\tilde{t}\equiv t/t_{RC}$. It is expressed in a lesser known form relating the radius to its second derivative:
\begin{equation}
\ddot{r}=-\tau^{2}r^{-4}
\label{eq:obr1}
\end{equation} 
The normalized equation is generalized for arbitrary dimension by differentiating (\ref{eq:drdt}) with respect to $t$:
\begin{equation}
\ddot{r}=-(N-2)\tau^{2}r^{-(N+1)}
\label{eq:obr1b}
\end{equation} 
This allows an examination of the behavior of the collapsing radius for various dimensions independent of timescale. It can be seen that for higher dimensions, the bulk of the collapse occurs towards the end of the process, while at lower dimensions it is a more gradual process (Figure \ref{fig:bub1}). This is noted by Prosperetti \cite{ProspBub}, who states that the order of the singularity at $\tilde{t}=1$ increases with dimensionality. To approximate $r(\tilde{t})$, Obreschkow et al. make an ansatz based on the general behavior of the collapse:
\begin{equation}
r_{o}=(1-\tilde{t}^{2})^{\alpha}
\label{eq:obr2}
\end{equation}
The exponent $\alpha$ is determined based on the asymptotic behavior of the bubble, and is derived in two ways. One can match the second derivative of $r(\tilde{t}=0)$ to $r_{o}(\tilde{t}=0)$, which yields an exponent of $\tau^{2}/2\approx0.418$, or one can match the behavior of the first derivative at $\tilde{t}=1$, which yields an exponent of $2/5=0.4$. Obreschkow et al. use the latter, and calculate a series of higher-order corrections to find better agreement between $r$ and $r_{o}$.

The value of $\alpha$ from the second derivative can be found by equating (\ref{eq:obr1b}) with the second derivative of (\ref{eq:obr2}), which yields the exponent:

\begin{equation}
\alpha=\tau^{2}\frac{N-2}{2}=\frac{\pi}{4N}\frac{\Gamma\left(\frac{N+2}{2N}\right)^{2}}{\Gamma\left(\frac{N+1}{N}\right)^{2}}\approx \frac{\pi^2}{4N}
\label{eq:alpha1}
\end{equation}
For large $N$ the gamma functions reduce to a factor of $\pi$. The other method, based on the first derivative at the time of collapse, expresses the first derivative in terms of the radius: $\dot{r_o}=r_{o}^{1-1/\alpha}$. This can be equated to (\ref{eq:drdt}) with $R_{o}=1$ in the limit $R\rightarrow 0$. This yields the other value for the exponent:
\begin{equation}
\alpha=\frac{2}{N+2}
\label{eq:alpha2}
\end{equation}
As with the three-dimensional case discussed by Obreschkow et al., the first method of calculating $\alpha$ finds better agreement with the full Rayleigh collapse equation at t=0 while the second method finds better agreement at t=1. Both methods, however, are within one-percent of the numerical solution until the end of the collapse. The two values of $\alpha$ are within five percent in three dimensions; in the large-$N$ limit the ratio approaches $\frac{\pi^{2}}{8}\approx 1.23$. The three-dimensional case finds the best agreement between the two values, although the agreement is exact if the two-dimensional case is allowed. Obreschkow et al. chose to focus on the second definition of $\alpha$, and insight from higher dimensions suggests that they chose correctly: the dynamics at late times in higher dimensions necessitate better accuracy in this limit rather than at the onset of collapse. A better understanding of the velocity at its zenith is necessary for studies of cavitation damage, one of the motivations of bubble research. Subsequent to the derivation of their approximation, Obreschkow et al. found a series of polynomial coefficients to improve the precision of their approximation, and from the scaling of these coefficients found a logarithmic integral that was true to the exact result. These coefficients and their scaling can be derived for arbitrary dimension, but this will be relegated to future work.

\subsection*{The Rayleigh-Plesset Equation}

The Rayleigh-Plesset in three dimensions describes the oscillations of a gas-filled spherical bubble. Its derivation is discussed in detail by Leighton \cite{leighton2}, both in terms of energy balance methods and the Navier-Stokes equation. It is a nonlinear second-order ordinary differential equation that takes the following form:
\begin{equation}
R\ddot{R}+\frac{3}{2}\dot{R}^2=\frac{1}{\rho}\left[P_o\left(\frac{R_o}{R}\right)^{3\gamma}-P_{\mathrm{ext}}\right]
\label{eq:RP3}
\end{equation}
The external pressure can include driving terms, such as an applied periodic acoustic field, as used in simulations of ultrasound contrast agents \cite{klotz2}. The $N$-dimensional analogue of the Rayleigh-Plesset equation can be derived assuming zero mass transfer, demanding the pressure in the bubble be a function of its radius, $P(R)$. The work performed by the bubble during oscillation is:
\begin{equation}
W=\int_{R_{o}}^{R}S_{N}R^N\left[P(R)-P_{ext}\right]dR
\label{eq:RPwork}
\end{equation}
By equating the kinetic energy (\ref{eq:energy}) and work (\ref{eq:RPwork}) and differentiating, the following relationship emerges:
\[\frac{S_{N}\rho}{2}\frac{\dot{R}^{2}R^{N}}{N-2}=\int_{R_{o}}^{R}S_{N}R^N\left[P(R)-P_{\mathrm{ext}}\right]dR \]
\[\frac{S_{N}\rho}{2(N-2)}\left[N\dot{R}^{2}R^{N-1}+2R^{N}\ddot{R}\right]=S_{N}R^N\left[P(R)-P_{\mathrm{ext}}\right] \]
\begin{equation}
\frac{R\ddot{R}}{N-2}+\frac{N\dot{R}^{2}}{2(N-2)}=\frac{P(R)-P_{\mathrm{ext}}}{\rho}
\label{eq:work1}
\end{equation}
Which is the analogue of the Rayleigh-Plesset equation in $N$ dimensions. We assume an ideal polytropic gas with the following pressure-volume relationship:
\begin{equation}
P_{o}V_{o}=P(R)V^{\gamma} \rightarrow P(R)=P_{o}\left(\frac{R_o}{R}\right)^{N\gamma}
\label{eq:polytropic}
\end{equation}
Gamma is the polytropic exponent, or ratio of specific heats. For an isothermal process, $\gamma$ is equal to one. For adiabatic systems, it is equal to $\frac{f+2}{f}$, where $f$ is the number of degrees of freedom. For a monatomic ideal gas with $N$ translational degrees of freedom, $f=N$ and $\gamma=\frac{N+2}{N}$. In the large-$N$ limit, the difference between the adiabatic and isothermal cases vanish.
Without surface tension or viscous damping, and assuming the equilibrium and external pressures are equal, the full Rayleigh-Plesset analogue is:
\begin{equation}
\frac{R\ddot{R}}{N-2}+\frac{N\dot{R}^{2}}{2(N-2)}=\frac{P_o}{\rho}\left[\left(\frac{R_o}{R}\right)^{N\gamma}-1\right]
\label{eq:ndrp}
\end{equation}
If the equilibrium radius is zero, this becomes a re-expression of Rayleigh-like collapse. The dynamics of $N$-dimensional bubbles were also discussed by Prosperetti \cite{ProspBub}, who derives similar expressions in a review of cavitation science. There is disagreement between the equations presented by Prosperetti and those of this paper. For example, the $N$-dimensional Rayleigh-Plesset equation according to Prosperetti merely replaces 3/2, the coefficient of the velocity term, with N/2. Obviously, experiments cannot be performed in higher dimensions to resolve this discrepancy, but the author finds fault in Prosperetti's derivation: his equation 3 does not follow from his equation 1, as it drops a dimensionally dependent factor from the second derivative. Other aspects of the Rayleigh-Plesset equation in higher dimensions, including the effects of surface tension and viscosity, are correctly discussed by Prosperetti.

There is a case in which the Rayleigh-Plesset equation is solvable: that of inertial expansion when the pressure term vanishes. Its solution is a power law in time with the same exponent as (\ref{eq:alpha2}), which was also derived by examining inertial behavior.
\begin{equation}
\frac{R\ddot{R}}{N-2}+\frac{N\dot{R}^{2}}{2(N-2)}=0 \rightarrow R(t)=R_{o}t^{\frac{N}{N+2}}
\label{eq:inertial}
\end{equation}

\subsection*{The Minnaert Frequency}

The Minnaert frequency is the resonance frequency of a spherical bubble at low amplitudes. It is derived by treating the bubble as a harmonic oscillator and takes the following form in three dimensions:
\begin{equation}
\omega_{M}=\sqrt{\frac{3\gamma P}{\rho R^{2}_{o}}}
\label{eq:minn}
\end{equation}
The Minnaert frequency in $N$ dimensions is derived by following a similar method as is used for three, again assuming small-amplitude oscillations and treating the system as a simple harmonic oscillator. The effective mass can be calculated by equating the kinetic energy (\ref{eq:energy}) with the classical definition, $\frac{1}{2}mv^2$, where the velocity is that of the bubble wall:
\begin{equation}
M_{\mathrm{eff}}=S_{N}\rho\frac{R^N}{N-2}
\label{eq:Meff}
\end{equation}
The effective spring constant can be found by assuming $R=R_o+\epsilon(t)$, where $\epsilon<<R_o$, and substituting it into the work integral with a first-order Taylor expansion, and extracting the spring constant assuming a harmonic potential: $W=\frac{1}{2}k\epsilon^{2}$.
	\[	W=\int_{0}^{\epsilon}P_{o}S_{N}R^{N-1}\left[\left(\frac{R_o}{R_o+\epsilon}\right)^{N\gamma}-1 \right]d\epsilon
\]
	\[	=\int_{0}^{\epsilon}P_{o}S_{N}R^{N-1}\left[\left(1+\frac{\epsilon}{R_o}\right)^{-N\gamma}-1 \right]d\epsilon 
\]	
	\[\approx\int_{0}^{\epsilon}P_{o}S_{N}R^{N-1}\left[\left(1-N\gamma\frac{\epsilon}{R_o}\right)-1\right]d\epsilon 
\]
	\[	=\int_{0}^{\epsilon}P_{o}S_{N}R^{N-1}\left(-N\gamma\frac{\epsilon}{R_o}\right)d\epsilon
\]
Evaluating:
\begin{equation}
W=\frac{P_{o}S_{N}N\gamma}{2}\epsilon^{2}R_{o}^{N-2}
\label{eq:Work}
\end{equation}
Extracting the spring constant we find:
\begin{equation}
k_{\mathrm{eff}}=S_{N}P_{o}N\gamma R_{o}^{N-2}
\label{eq:k}
\end{equation}
The Minnaert angular frequency, $\omega_N$ represents the natural frequency of low amplitude oscillations in $N$ dimensions:
\begin{equation}
\omega_N=\sqrt{\frac{k_{\mathrm{eff}}}{M_{\mathrm{eff}}}}=\sqrt{\frac{N(N-2)\gamma P_o}{\rho R_{o}^2}}
\label{eq:minnaert}
\end{equation}
The resonance frequency increases with increasing dimension, approaching a linear relationship for high dimension. Like the collapse scenario, the dynamics are faster in higher dimensions. One may note that in the isothermal case ($\gamma=1$), the Rayleigh collapse time $\tau$ in three dimensions is close to one-quarter of the Minnaert period: in dimensionless form, $\tau\approx0.914$, $2\pi/4\omega\approx0.907$. Simply stated, the time it takes for a perturbed gaseous bubble to return to its equilibrium radius is similar to the time it takes for a vacuum bubble of the same radius to collapse, to within one-percent. This is, however, unique to three-dimensions: the ratio of the bubble collapse time and perturbation can be found by dividing (\ref{eq:trc}) by (\ref{eq:minnaert}). In the large $N$ limit this ratio approaches the square root of two.

Having derived the chief results for bubble collapse and oscillation in higher dimensions, the general behavior of these systems can be examined.

\begin{figure}[ht]
	\centering
		\includegraphics[width=0.45\textwidth]{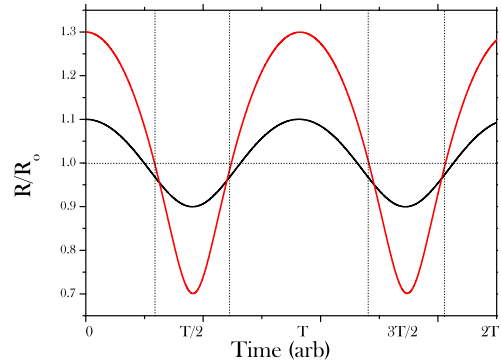}
	\caption{The onset of nonlinear behavior. With larger amplitudes, the bubble radius spends less time below equilibrium and the period begins to shift.}
	\label{fig:bub2}
\end{figure}

\section{Discussion: The Onset of Nonlinear Behavior}

While the total parameter space in which the Rayleigh-Plesset equation can be simulated is too vast for a single paper, a simple case can be examined to study the onset of nonlinear behavior. A dimensionless bubble is studied by numerically integrating (\ref{eq:ndrp}) with varying initial conditions.

The Rayleigh-Plesset equation at low amplitudes describes a harmonic oscillator. At larger amplutudes, the oscillations approach those of a cycloid. In the regime $R>>R_o$ the internal pressure becomes insignificant and the dynamics can be described by (\ref{eq:drdt}) which is of similar form, to the equation of a cycloid. This non-sinusoidal breathing mode is characterized by a peak wall velocity that is much larger near the minimum radius than near the maximum, and the oscillations are asymmetric about the equilibrium radius, spending much less time at small radius (Figure \ref{fig:bub2}). In addition to the waveform changing from sinusoidal to cycloidal, there is also an upward shift (in three dimensions) of the oscillation period of the bubble, with the shift increasing with amplitude. This is analogous to the simple pendulum in the nonlinear regime, where the period is dependent on the initial amplitude. This nonlinear effect has been observed experimentally by driving bubble oscillations with ultrasound \cite{Lauterborn}.

The onset of nonlinearity was investigated in higher dimensions by simulating numerically (\ref{eq:ndrp}) with and calculating the relative radius $R(t)/R_o$ as a function of time. The initial velocity was zero and the initial radius was varied from below to above the equilibrium value. This defined the independent variable, the initial strain $R(0)/R_o$. The nonlinearity can be quantified in several ways: the fraction of the oscillation the bubble spends below its equilibrium radius, the appearance of harmonics in a Fast Fourier Transform of the waveform, or the shift in the oscillation period or frequency. The period shift is primarily is used, as it can be measured from a simulation most precisely without dependence on numerical parameters and highlights several interesting features.

\begin{figure}[ht]
	\centering
		\includegraphics[width=0.45\textwidth]{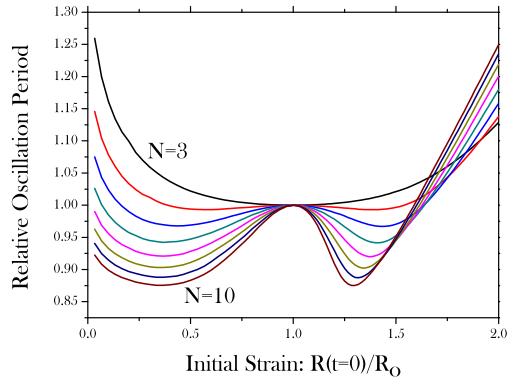}
	\caption{The shift in oscillation period as a function of initial strain for dimensions 3-10. Dimensions greater than three have non-monotonic behavior with  a maximal period shift, while the nonlinear behavior is stronger at higher dimensions.}
	\label{fig:bub3}
\end{figure}

The period as a function of initial strain at various dimensions can be seen in Figure \ref{fig:bub3}. In the case where the initial strain is greater than one, the oscillation period increases with increasing strain in three dimensions. In higher dimensions, it decreases and reaches a local minimum, before increasing monotonically after that minimal point. The onset of this behavior as a function of initial strain was greater in higher dimensions. For larger initial strains, the period increased towards infinity, with a greater rate of increase in higher dimensions. The period shift approaches a linear relationship with initial radius, which is expected when the bubble's oscillations approximate Rayleigh-like collapse. The effect on the period was symmetric about $\frac{R(0)}{R_o}=1$ for small strains, but asymmetry arose due to the fact that the initial size of the bubble cannot be less than zero. It is believed that in the limit of the initial radius going to zero, the period of oscillation doubles. However, computational resources were only available to confirm this up to $N=6$. This limit does not exist for bubbles with initial radii greater than $R_o$.

\begin{figure}[ht]
	\centering
		\includegraphics[width=0.45\textwidth]{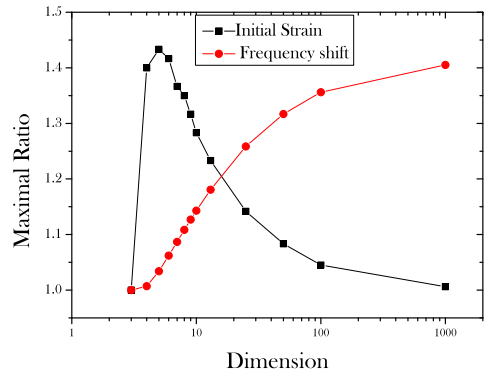}
	\caption{The initial strain that leads to a local maximum in the oscillation frequency (or minimum in the period, Figure {fig:bub3}), and the maximal frequency at that strain, as a function of dimension. It occurs with the greatest strain at $N=5$, and the relative frequency trends towards the square root of two.}
	\label{fig:bub4}
\end{figure}

The initial strain at which the greatest positive frequency shift (or negative period shift) occurs was greatest in five dimensions (Figure \ref{fig:bub4}). The strain that yielded the greatest positive frequency shift occurred approached unity with increasing dimension, and the maximal positive frequency shift approached the square root of two relative to the Minnaert frequency. This ratio, however, approaches the square root of two differently than does the ratio of Rayleigh collapse time to Minnaert relaxation time.

\begin{figure}[ht]
	\centering
		\includegraphics[width=0.45\textwidth]{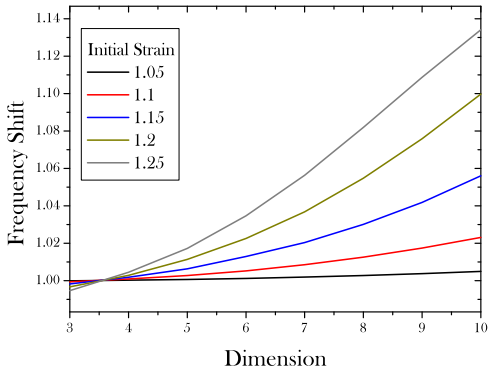}
	\caption{Frequency shift as a function of dimension for various initial strains. Nonlinear behavior is stronger at higher dimensions.}
	\label{fig:bub5}
\end{figure}

These simulations also allow the two-dimensional case to be examined by considering the limiting behavior of a $2+\epsilon$ dimensional bubble. While the period of oscillations diverges as $\epsilon\rightarrow 0$, the shift in frequency is largely independent when $\epsilon$ is small: the frequency shift follows the same behavior as the three-dimensional bubble, without a local maximum. To quantify the onset of nonlinearity as a function of dimensionality, the frequency shift as a function of dimensionality for a given initial strain can be examined (Figure \ref{fig:bub5}). The frequency shift at a given strain increases monotonically with dimension, with the rate of increase dependent on the initial strain. To characterize the transition from sinusoidal to cycloidal oscillations, the fraction of the period spent below equilibrium was examined, and showed similar trends to the frequency shift with respect to dimensionality: nonlinear behavior is stronger in higher dimensions. This, as well as the dimensional dependence of the period shift, highlights the fact that the onset of nonlinearity occurs more readily in higher dimension. These simulations highlight a few features unique to three dimensions: the three-dimensional bubble maintains a sinusoidal breathing mode at the greatest amplitudes, and the amplitude-dependence of frequency is only monotonic. At higher dimensions, there exists a certain large amplitude at which the cycloidal oscillations have the same frequency as the low-amplitude sinusoidal oscillations, but not in three. This behavior is not fully understood and merely serves to characterize the nonlinear behavior of bubble oscillations in higher dimensions. Because the behavior near the minimum size involves wall velocities that approach or exceed that of sound in the fluid, a model that takes into account compressibility may be necessary to fully characterize and understand this behavior.

\section{Concluding Remarks}

Several canonical results in bubble dynamics, including the Rayleigh collapse time, the Rayleigh-Plesset equation, and the Minnaert frequency have been re-derived in arbitrary dimension. The bubbles generally display faster dynamics in higher dimensions, as the natural period and collapse time both decrease. Particularly, the late-time collapse behavior of a bubble is much stronger in higher dimensions. The normalized collapse equation and its analytic approximation were examined in higher dimensions, suggesting a choice of boundary conditions upon which to base the approximations. 
The linear and nonlinear oscillations of a bubble were simulated numerically with respect to the initial strain of the bubble's radius. It was observed that in higher dimensions the onset of nonlinear behavior, characterized by cycloidal oscillations with varying frequency, occurs at lesser amplitudes.

Many factors were not discussed in this paper, including the effect of fluid compressibility and viscosity. Because the dynamics in higher dimensions are faster, the derivation of a compressible fluid model of $N$-dimensional bubble oscillations presents itself as a follow-up. While the generalized derivations used in this paper do not apply to one- and two-dimensional systems, their quasi-low-dimensional real-world analogues have been discussed in detail elsewhere \cite{sass}. Several features were found to be particular to three dimensions: the collapse time of an empty bubble and the relaxation time of an isothermal gas-filled bubble are nearly the same only in three dimensions, and the two dynamical exponents of the analytic collapse approximation are closest in three dimensions. The three-dimensional bubble exhibits nonlinear behavior, but this behavior is stronger in higher dimensions. The shift of frequency with respect to amplitude is only monotonic in three dimensions, while bubbles in higher dimensions display non-monotonic behavior, having a range of strains that lead to an upward shift in frequency. 

Lacking from this paper are true attempts to use the higher-dimensional information to better understand three-dimensional bubble behavior or solve the relevant equations. Solution schemes to the Rayleigh-Plesset equation based on perturbation analysis \cite{perturb} can be examined in the light of the $N$-dimensional Rayleigh-Plesset equation to improve upon those methods. These derivations may be coupled to higher dimensional models of chemical or plasma physics to better understand sonoluminescence. It is hoped that in the future more insight into three-dimensional bubble dynamics can be gained by examining their behavior in higher dimensions.

\bibliographystyle{unsrt}
\bibliography{bubblereferences}

\end{document}